\documentclass[aps,prb,a4paper,twocolumn,superscriptaddress,floatfix]{revtex4-2}

\usepackage{amssymb,amsmath,amsfonts}
\usepackage{graphicx}
\usepackage[dvipsnames]{xcolor}
\usepackage{hyperref}
\usepackage{bm}
\usepackage{comment}
\usepackage{cancel}
\usepackage{subcaption}

\DeclareMathOperator{\re}{Re}

\DeclareMathOperator{\curl}{curl}
\DeclareMathOperator{\dive}{div}
\DeclareMathOperator{\grad}{grad}

\def\bA{\mathbf{A}}
\def\br{\mathbf{r}}

\def\bj{\mathbf{j}}
\def\bE{\mathbf{E}}
\def\bH{\mathbf{H}}
\def\bB{\mathbf{B}}
\def\bD{\mathbf{D}}

\def\be{\begin{equation}}
\def\ee{\end{equation}}
\def\bea{\begin{eqnarray}}
\def\eea{\end{eqnarray}}
\def\bc{\begin{cases}}
\def\ec{\end{cases}}

\begin{document}

\title{Superconducting photocurrents induced by structured electromagnetic radiation}
\author{O.~B.\ Zuev}
\affiliation{Moscow Institute of Physics and Technology (National research university), Dolgoprudny, Moscow region 141700, Russia} 
\affiliation{L.~D.\ Landau Institute for Theoretical Physics,  Chernogolovka 142432, Russia}

\author{M.~V. Kovalenko}
\affiliation{Moscow Institute of Physics and Technology (National research university), Dolgoprudny, Moscow region 141700, Russia} 
\affiliation{L.~D.\ Landau Institute for Theoretical Physics,  Chernogolovka 142432, Russia}

\author{A.~S.\ Mel'nikov}
\affiliation{Moscow Institute of Physics and Technology (National research university), Dolgoprudny, Moscow region 141700, Russia} 
\affiliation{Institute for Physics of Microstructures, Russian Academy of Sciences, 603950 Nizhny Novgorod, GSP-105, Russia}

\date{\today}

\begin{abstract}
We develop a phenomenological theory describing the interaction of superconducting condensate with a Bessel beam of twisted light characterized by a nonzero angular momentum $m$. Starting from the time-dependent Ginzburg-Landau model with the complex relaxation time we calculate the spatial profiles of dc photoinduced currents and magnetic fields as well as the second harmonic response. The photocurrents and magnetic fileds are shown to be determined both by the helicity of light and its orbital momentum $m$. Analyzing the half-space and thin film geometries we discuss possible experimental tests aimed to probe the superconducting photocurrents and magnetic fields.

\end{abstract}

\maketitle

\section{\label{sec:one}Introduction}

Recent progress in the study of a variety of optoelectronic phenomena and their applications triggered the interest 
to the dissipationless photocurrents which can be generated in systems with superconducting elements. 
The key distinctive feature of these photocurrents carried by the Cooper pairs is that they can persist even after 
the switching off the electromagnetic pulse responsible for the current generation. Such persistent current 
becomes possible due to the photoinduced superconducting vortices penetrating the system acquiring, thus, the angular momentum
from the incident electromagnetic wave \cite{Plastovets22, Croitoru22}. The primary angular momentum of the incident light pulse can originate either from its circular polarization or from a nontrivial orbital structure of the light beam. The transfer of the angular momentum for the case of circularly polarized radiation can be illustrated even for the case of the plane wave and this phenomenon is well known as the inverse Faraday effect \cite{Pitaevskii61}. In the context of superconducting materials the mechanisms of this effect have been recently discussed in \cite{Buzdin21, Dzero24}. Turning to the case of the electromagnetic radiation with a nonzero orbital momentum we need to consider the so-called twisted light which can be realized in Bessel or Laguerre-Gaussian beams \cite{Allen92,Forbes21, Allen99, Molina-Terriza07} (see also \cite{Knyasev18} for review).    
Advances in the development and manipulation of the structured light beams have significantly expanded the possibilities for studying light-matter interactions. 
The propagation and scattering of beams in various media are of particular  interest as the study of these phenomena opens the possibility of deeper understanding of the physics of light-matter interaction (see, e.g., \cite{Ji20,Sederberg20}).
Note that the study of photogalvanic effects in media irradiated by structured light is also of practical interest, particularly, for the development of compact devices that enhance optical communication and data processing by utilizing the orbital angular momentum of light \cite{Andrewsbook}.

Considering the nonlinear electrodynamics of these beams we will focus only on the second-order nonlinear effects as they define the rectification phenomena. 
Typical second-order nonlinear effects occurring in metals, semiconductors, or plasma during interaction with radiation include the inverse Faraday effect \cite{Pitaevskii61, Hertel06, Karpman81}, the Gaponov-Miller force \cite{Gaponov-Miller, Hertel06, Karpman81, Perel73}, ac Hall effect \cite{Barlow58}, and related photon drag phenomenon \cite{Strait19, Glazov13, Normantas85, Durnev21, Gurevich93, Gurevich94, Ivchenko12}.
Detailed study of these effects for the particular case of structured light (i.e. light beams) illuminating semiconducting samples have been recently presented in Refs.~\cite{Gunyaga23,Gunyaga24}.

Turning now to the case of superconductors we have to note that for the plane waves the above second-order effects have been recently addressed in a number of works based mainly on the time - dependent Ginzburg - Landau theory. In particular, above the critical temperature $T_c$ in the fluctuating regime the second order nonlinearities  have been studied in Refs.~\cite{Boev20, Parafilo22, Boev22, Plastovets23}. 
Below $T_c$ the mechanisms responsible for the inverse Faraday effect and photon drag of Cooper pairs have been revealed in Refs.~\cite{Mironov24, Buzdin21}. These mechanisms originate from the time - dependent modulation of the superfluid density caused by the
charge imbalance potential $\tilde{\phi}$ induced by the field of the incident electromagnetic wave. As a result, the supercurrent contains the term proportional to the product of the superfluid velocity and charge imbalance potential generating, thus, the second - order response.

\begin{figure}
\centerline{\includegraphics[width=1.0\columnwidth]{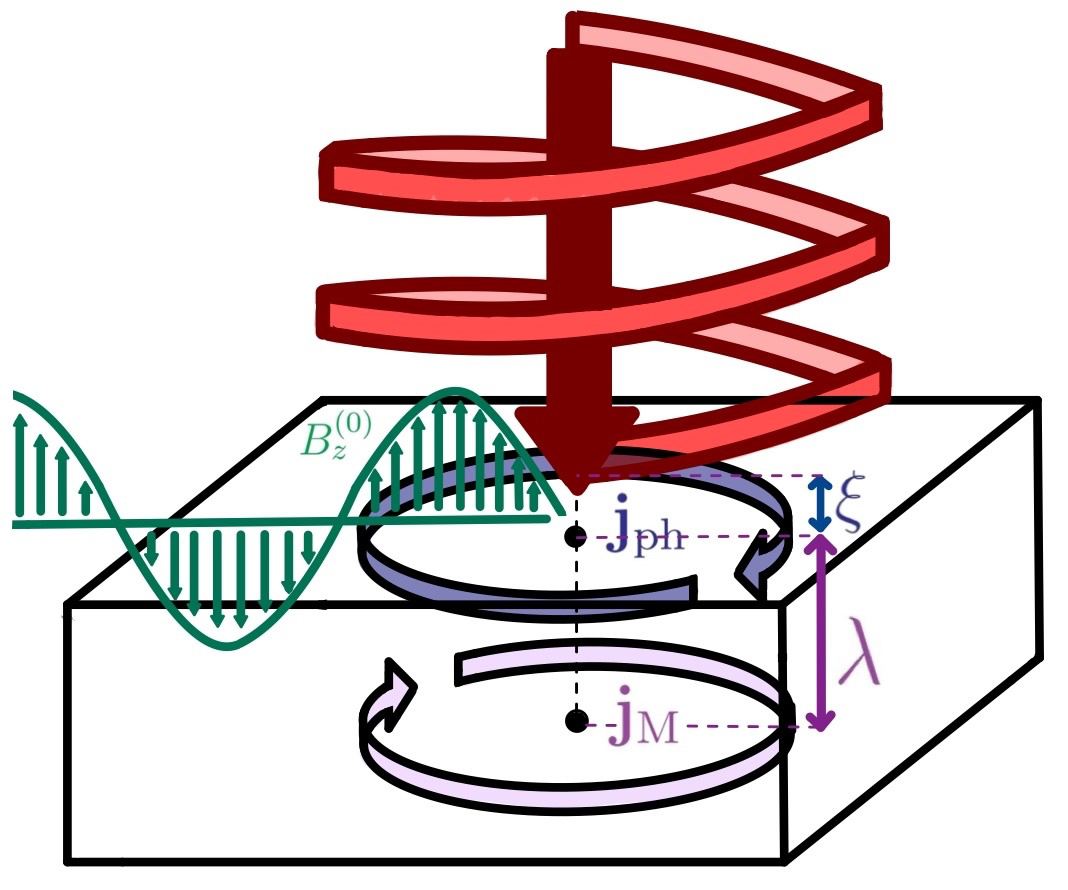}}
\caption{Schematic experimental setup for detection of photocurrents induced by structured light. The figure shows a beam incident on a superconductor, the photoinduced current $\mathbf{j}_\text{ph}$ flowing at the depth $\sim\xi$, and the screening Meissner current $\mathbf{j}_\text{M}$ flowing at the depth $\sim\lambda$.}
\label{Setup}
\end{figure}

The goal of the present work is to generalize the studies \cite{Mironov24, Buzdin21} to the case of structured light illuminating the superconducting condensate focusing on the particular case of Bessel beams.
The primary calculation tool in our work is the time-dependent Ginzburg-Landau (TDGL) formalism.
 The key ingredient which allows to study the second - order nonlinearities within this formalism is the imaginary part of the relaxation constant which is defined by the electron-hole asymmetry \cite{Aaronov92, Kopninbook}.
Note that accounting for the second order nonlinear response of the condensate and related rectification effects distinguishes our consideration from the previous studies of interaction of structured light with superconducting systems \cite{balatsky1,balatsky2}.
Following the approach developed in \cite{Mironov24} we calculate the magnitudes of the charge imbalance potential $\tilde{\phi}$ and order parameter modulation $\Delta_1$ in superconductor. We calculate the induced photocurrent and compensating Meissner currents  both at zero frequency and the second harmonic, and finally obtain the photoinduced magnetic field profiles  which can be measured by a certain local probe techniques [see Fig.~\ref{Setup}].
Based on our calculations for a superconducting half-space, the magnetic field at the interface at the center of the beam can be estimated as follows: $B_z \sim \nu \kappa^2\lambda\xi \frac{B_0^2}{H_\text{cm}}$,
where $\nu$ is the ratio between imaginary and real parts of the relaxation constant $\Gamma$ in TDGL model, $1/\kappa = c/(\omega\sin\theta)$ is the beam length scale in the sample plane, $\lambda$ is the London penetration depth, $\xi$ is the superconducting coherence length, $B_0$ is the characteristic amplitude of the magnetic field in the beam and $H_{cm}$ is the thermodynamic critical field of the superconductor. 
The total photoinduced magnetic field has two contributions: the one of orbital origin proportional to the  $m$ value and, thus, antisymmetric under the substitution $m \rightarrow -m$, and the one associated with the polarization rotation related, therefore, to the inverse Faraday effect. It is important to note that the  photocurrents and resulting magnetic fields can be strengthened in thin film geometry in the factor $(\lambda/d)^2$, if the film thickness $d$ is less than the London penetration depth $\lambda$.
Thus, the use of thin superconducting films in the experimental setup can facilitate the observation of the above effects.

The remainder of the paper is organized as follows. In Section \ref{sec:two} we give the basic equations of the time-dependent Ginzburg-Landau theory. In Section \ref{sec:three} we discuss the auxilliary problem of photon drag for the case of plane waves with mixed polarization. In Sections \ref{sec:four} and \ref{sec:five} we present the calculations of photocurrents and related magnetic fields for the Bessel beam interacting with superconducting half-space and thin film, respectively.
Finally, the results are summarized in Sec. \ref{conclusion}.

\section{\label{sec:two}Basic equations}

In our study of the second-order nonlinear electrodynamic response of  superconductor we follow the
approach adopted previously in \cite{Buzdin21, Mironov24, Plastovets23}, namely, 
 the time-dependent Ginzburg-Landau (TDGL) theory with a complex relaxation constant $\Gamma$:
\begin{multline}
\label{GL-equation}
-\Gamma\left(\hbar\partial_t - 2ie\phi\right)\Delta =
\\
- a T_c \epsilon\Delta + b |\Delta|^2\Delta + a T_c \xi_0^2 \left(-i \mathbf{\nabla} + \frac{2\pi}{\Phi_0} \mathbf{A} \right)^2\Delta
\end{multline}
where $ \Gamma = \frac{\pi a}{8} + i\gamma $, $\Phi_0=\pi\hbar c/e$ ($e>0$). The imaginary part of the relaxation constant is known to reflect a small electron-hole asymmetry.  Assuming that the external field is sufficiently weak, we adopt a perturbation expansion for the order parameter in the form $ \Delta = (\Delta_0 + \Delta_1)e^{i\chi} $, where $ \Delta_0^2 = a T_c \epsilon/b $ is the equilibrium order parameter in the absence of the external field, and $ \epsilon = (T_c - T)/T_c $. Corrections to the order parameter amplitude $ \Delta_1$, and phase $ \chi$ are assumed to be proportional to the amplitude of the electromagnetic wave.

Though the above model has a rather
restricted range of validity and assumes a gapless superconducting
state, such consideration is known to provide
instructive insights for a great variety of dynamic phenomena. Microscopic derivation of the TDGL model has been previously obtained for two possible scenarios when the gap suppresion is caused  either by paramagnetic impurities or by the presence of rather strong electron-phonon relaxation (see \cite{Kopninbook,Larkinbook}  for review).

Taking a single Fourier harmonic of the incident electromagnetic wave ($\propto e^{-i\omega t}$) one can write 
the second-order response of the superconductor through the complex amplitudes of the correction to the gap $\Delta_1$ and the superfluid velocity $\mathbf{v}_s$:
\begin{equation}
\label{jphdef}
\bj_\text{ph} = \bj_\text{ph}^{(0)} +  \bj_\text{ph}^{(2\omega)} = -2e\Delta_0 \re [
\left( \Delta^*_1 + \Delta_1 e^{-2i\omega t}\right) \mathbf{v}_s ]
\end{equation}

The equation for the correction $\Delta_1$ reads (see Appendix \ref{Appendix1} for details):
\begin{equation}
- \nabla^2 \Delta_1 + q_1^2 \Delta_1 = \frac{2e\tau \nu \Delta_0}{\hbar\xi^2}  \tilde{\phi}, \label{deltaeq} 
\end{equation}
where $ q_1^2 \equiv \frac{2-i\omega\tau}{\xi^2}$, $\xi^2 = \xi_0^2/\epsilon$, $\tau = \pi \hbar / 8 T_c \epsilon $, and the parameter $\nu=8\gamma/\pi a$ is the ratio of the imaginary and real parts of $\Gamma$.
Here we introduce a charge imbalance potential $\tilde{\phi} = \phi - \mu_p$, which is known to characterize the difference between the chemical potential of normal quasiparticles in a superconductor and the chemical potential of Cooper pairs $\mu_p = (\hbar/2e)\partial\chi/\partial t$ \cite{Artemenko79}.
The equation for the charge imbalance potential $\tilde{\phi}(\br)$ takes the form (see Appendix \ref{Appendix1} for details):
\begin{equation}
\nabla^2 \tilde{\phi} = q_2^2 \tilde{\phi}, \label{phieq} 
\end{equation}
where $ q_2^2 \equiv \frac{1}{l_E^2} - \frac{i\omega\tau}{\xi^2} = \frac{\lambda^2}{l_E^2 \lambda_\text{eff}^2}$, $l_E^2 = \hbar \sigma_n / \pi a e^2 \Delta_0^2$, $\lambda^2 = \hbar^2 c^2 / 32\pi a T_c \xi_0^2 e^2 \Delta_0^2$, $\frac{1}{\lambda_\text{eff}^2} = \frac{1}{\lambda^2} \left(1-i\omega\tau \frac{l_E^2}{\xi^2}\right)$. Note that for further estimates we will take $l_E = \xi/\sqrt{5.79}$ keeping in mind the parameters of the TDGL model derived for the case of strong electron-phonon relaxation (see \cite{IvlevKopnin}).

The superfluid velocity in the expression \eqref{jphdef} can be obtained from the Maxwell's equations.  
In the linear approximation the magnetic and electric fields inside superconductor ($\mathbf{H}^{(i)} $ and $\bE^{(i)}$) satisfy 
the following equations derived in Appendix \ref{Appendix2}:
\begin{equation}
\label{Hequation}
\Delta \mathbf{H}^{(i)} = \frac{1}{\lambda_{\text{eff}}^2} \mathbf{H}^{(i)} \ ,
\end{equation}
\begin{equation}
\label{Eequation}
\curl\mathbf{H}^{(i)} = \frac{ic}{\omega} \left(\frac{1}{\lambda_\text{eff}^2} \bE^{(i)} + \frac{1}{\lambda^2} \nabla\tilde{\phi}\right) \ .
\end{equation}

The superfluid velocity can be expressed through the field $\bE^{(i)}$: 
\begin{multline}
\label{vsdef}
\mathbf{v}_s = -\frac{4i e a T_c \xi_0^2}{\omega \hbar^2}\left(\bE^{(i)} + \nabla\tilde{\phi}\right) \approx -\frac{4i e a T_c \xi_0^2}{\omega \hbar^2} \bE^{(i)} \approx 
\\
-\frac{4e a T_c \xi_0^2 \lambda^2_\text{eff}}{c \hbar^2}\curl \bH^{(i)} \ .    
\end{multline}
Here, the term $\nabla\tilde{\phi}$ appears to be small in the parameter $\omega/\sigma_n$ and can be omitted. Indeed, the use of the Leontovich boundary condition for the tangential components of the fields at the surface of metal assumes that the normal to the surface electric field component is small in the above ratio $\omega/\sigma_n$. It is this normal field component which generates the gradient of the charge imbalance potential inside the superconductor and, thus, its value is also determined by the small factor $\omega/\sigma_n$.
The upper limit for the frequencies considered in our work is obviously determined by the superconducting gap value and, thus, should be in the THz frequency range ($\omega \lesssim 10^{12} s^{-1}$). These values are certainly much less than typical values of conductivity  $\sigma_n \sim 10^{18} s^{-1}$.

\section{\label{sec:three}Auxilliary problem. Photon drag for mixed polarization of the incident wave}

\begin{figure}
\centerline{\includegraphics[width=1.0\columnwidth]{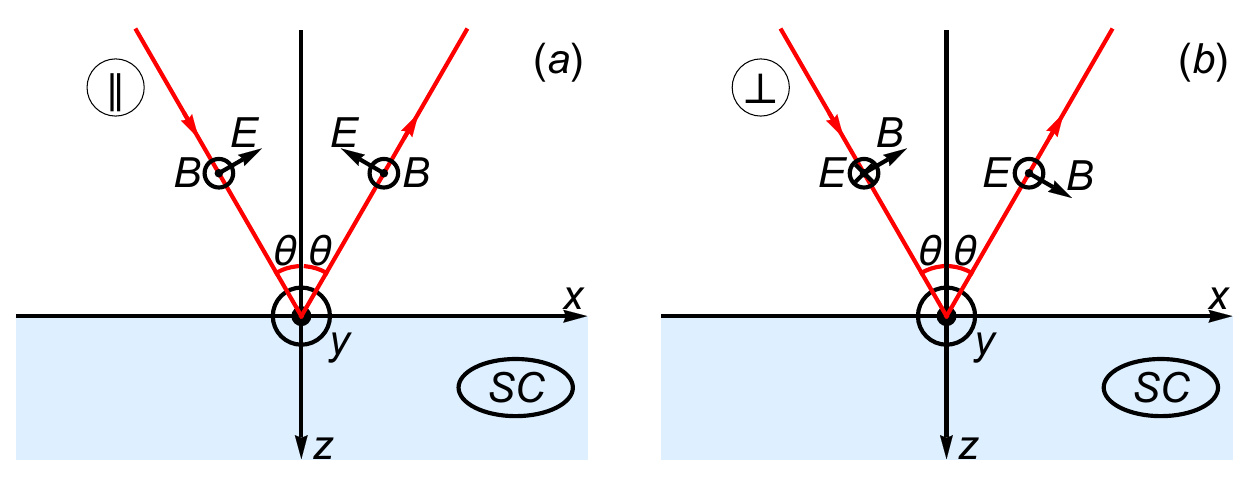}}
\caption{Incident and reflected wave structure for P-polarization (a) and for S-polarization (b). 
}
\label{PSpolarizations}
\end{figure}

Keeping in mind that further consideration of the structured light requires to take account  of the coexistence of the waves of different polarization it is convenient first to generalize the description of Ref. \cite{Mironov24} for the case of photon drag arising in the presence of electromagnetic wave with both P- and S- polarizations (the corresponding orientations of electric and magnetic fields are shown in Fig.\ref{PSpolarizations}). This auxilliary problem should illustrate how these polarizations mix on the second order nonlinearity in the superconductor response and create the dc current.
Thus, we consider the electromagnetic wave incident on a superconductor surface consisting of the mixture of P- and S- polarized parts. For P-polarization the complex amplitudes of  incident and reflected waves (omitting the factor $e^{-i\omega t}$) are given by the expressions (the unspecified components are equal to zero, see Fig.\ref{PSpolarizations}):
\begin{equation}
\label{P-fields}
\begin{cases}
E_x^{(e)} = E_0 \cos\theta e^{i k_x x} \left(e^{i k_z z} - r_\parallel e^{-i k_z z}\right) \ ,
\\
B_y^{(e)} = B_0 e^{i k_x x} \left(e^{i k_z z} + r_\parallel e^{-i k_z z}\right) \ ,
\\
E_z^{(e)} = -E_0 \sin\theta e^{i k_x x} \left(e^{i k_z z} + r_\parallel e^{-i k_z z}\right) \ .
\end{cases}
\end{equation}
For S-polarization these amplitudes take the form (see Fig.\ref{PSpolarizations}):
\begin{equation}
\label{S-fields}
\begin{cases}
B_x^{(e)} = B_0 \cos\theta e^{i k_x x} \left(e^{i k_z z} + r_\perp e^{-i k_z z}\right)
\\
E_y^{(e)} = -E_0 e^{i k_x x} \left(e^{i k_z z} - r_\perp e^{-i k_z z}\right)
\\
B_z^{(e)} = -B_0 \sin\theta e^{i k_x x} \left(e^{i k_z z} - r_\perp e^{-i k_z z}\right)
\end{cases}
\end{equation}
Hereafter indices $(e)$ and $(i)$ stand for the external and internal fields.
Next we consider the superposition of two polarizations and introduce 
the polarization unit vector corresponding to the magnetic field direction:
\begin{equation}
\mathbf{e}_\text{H} = \alpha \mathbf{e}_\text{H}^{\parallel} + \beta \mathbf{e}_\text{H}^{\perp} 
 \ ,
\end{equation}
where
\begin{equation}
\begin{cases}
\mathbf{e}_\text{H}^{\parallel} = (0, 1, 0)^T,
\\
\mathbf{e}_\text{H}^{\perp} = (\cos\theta, 0, -\sin\theta)^T.
\end{cases}
\end{equation}
The complex coefficients 
$\alpha$ and $\beta$) are the amplitudes of the P- and S- polarization contributions to the fields,
 $|\alpha|^2 + |\beta|^2 = 1$.

In order to calculate the reflection coefficients, we neglect the term $\nabla\tilde{\phi}$ in equation \eqref{Eequation} assuming the limit of $\omega \ll \sigma_n$ and $l_E \ll \lambda$, which looks reasonable for a II-type superconductor:
\begin{equation}
\label{Eeqreduced}
\curl\bH^{(i)} = \frac{ic}{\omega \lambda_\text{eff}^2} \bE^{(i)}. 
\end{equation}
Taking into account the expressions \eqref{P-fields} and \eqref{S-fields},
we obtain the reflection coefficients $r_\parallel$ and $r_\perp$ for P- and S- polarizations, correspondingly (see \cite{LL8}): 
\begin{equation} 
\label{r_parallel}
r_\parallel = \frac{1 + i\frac{\omega \lambda_\text{eff}}{c \cos{\theta}}}{1 - i\frac{\omega \lambda_\text{eff}}{c \cos{\theta}}} \ , 
\end{equation}
\begin{equation} 
\label{r_perp}
r_\perp = \frac{1 + i\frac{\omega \cos{\theta} \lambda_\text{eff}}{c}}{1 - i\frac{\omega \cos{\theta} \lambda_\text{eff}}{c}}. 
\end{equation}

Eq. \eqref{Hequation} gives us the solution for the magnetic field inside the superconductor. Exploiting the continuity of the tangential component of the magnetic field we obtain
\begin{equation}
H^{(i)}_y = (1+r_\parallel)\alpha B_0 e^{i k_x x-z/\lambda_\text{eff}}
\end{equation}
for P-polarization and 
\begin{equation}
H^{(i)}_x = (1+r_\perp)\beta B_0\cos\theta e^{i k_x x-z/\lambda_\text{eff}}.
\end{equation}
for S-polarization.
Hereafter we put the reflection coefficients $r_\parallel=1$ and $r_\perp = 1$ assuming that $\omega |\lambda_\text{eff}|/c \ll 1$ and taking $\theta$ not very close to $\pi/2$. Therefore, the $z$-component $H^{(i)}_z \propto (1-r_\parallel)$ is omitted. 

To compute the light-induced current, we need the correction to the absolute value of the order parameter $\Delta_1(\br)$ (see Eq.\eqref{jphdef}).  
Note that since the $z$-component of the electric field for S-polarization is zero (see Fig. \ref{PSpolarizations}), this polarization does not contribute to $\tilde{\phi}$ and $\Delta_1$: $\tilde{\phi}=\Delta_1=0$. Thus, we need to calculate the quantity $\Delta_1(\br)$ only for P-polarization.

Disregarding a very thin surface layer (of the thickness of Thomas-Fermi screening length $\lambda_\text{TF} \ll \xi$) accumulating the oscillating surface charge we can choose the boundary condition for the charge imbalance potential in the form 
\begin{equation}
\partial \tilde{\phi}/\partial z|_{z=0} = -E_z^{(i)}|_{z=0} = -\frac{c}{4\pi\sigma} \partial H_y^{(i)}/\partial x |_{z=0} \ .
\end{equation}
The corresponding solution of Eq.  \eqref{phieq} reads 
\begin{equation} 
\tilde{\phi} = \alpha\frac{i \omega \sin\theta B_0}{2\pi\sigma_n q_2} e^{ik_x x} e^{-q_2 z} \ . 
\end{equation}

Using the above expressions and the boundary condition $\partial \Delta_1/\partial z|_{z=0}=0$ we find the solution of Eq. \eqref{deltaeq}:
\begin{multline}
\Delta_1 = -\alpha\frac{i\omega e \tau \nu \Delta_0 B_0\sin\theta}{\pi\hbar\sigma_n \xi^2(q_1^2 - q_2^2)} 
e^{ik_x x}\left( \frac{e^{-q_1 z}}{q_1} - \frac{e^{-q_2 z}}{q_2} \right) .
\end{multline}

Neglecting the small corrections in the parameter $\omega \lambda/c$ we get the expression  
\begin{equation}
\curl \bH^{(i)} \approx \begin{pmatrix}
- \partial H^{(i)}_y/\partial z \\ \partial H^{(i)}_x/\partial z \\
\partial H^{(i)}_y/\partial x
\end{pmatrix}
= \frac{2B_0}{\lambda_\text{eff}}
\begin{pmatrix}
\alpha \\ -\beta \cos\theta \\
i\alpha \sin\theta \lambda_\text{eff} \omega/c 
\end{pmatrix} \ ,
\end{equation}
which allows us to find the superfluid velocity according to the Eq. \eqref{vsdef}. Here the column vectors are written in the Cartezian coordinate system. 

Finally, in the presence of both polarizations we get the rectified current density vector at zero frequency $\mathbf{I}^{(0)}_{\text{ph}} = -2e \Delta_0 \re\int_0^{+\infty} dz \Delta_1^* \mathbf{v_s}$:
\begin{multline}
\label{Izeroplane}
\mathbf{I}^{(0)}_{\text{ph}} = \frac{\nu}{\sqrt{2}\pi} \frac{\xi}{\tau} \frac{B_0^2\sin{\theta}}{H_{cm}} \frac{\omega \tau}{1+\omega^2\tau^2\eta^2} \times
\\
\re\left[ i \frac{\sqrt{1+i\omega\tau\eta}}{2-i\omega\tau} \begin{pmatrix}
|\alpha|^2 \\ -\alpha \beta^* \cos{\theta} \\ -i|\alpha|^2\sin\theta (\lambda\omega/c\sqrt{1+i\omega\tau\eta})
\end{pmatrix} \right] \ ,
\end{multline}
where the parameter $\eta = l_E^2/\xi^2$ characterizes the ratio between the superconducting coherence length and the electric field penetration depth. 
Note that in the limit $\alpha=1,\beta=0$ the Eq.\eqref{Izeroplane} gives us the result for P-polarization  obtained in \cite{Mironov24}. The $y$-component of the Eq. \eqref{Izeroplane}  proportional to $\alpha\beta^*$ is responsible for the mixing of polarizations due to second-order nonlinearity: the wave with S-polarization induces the in-plane screening current while the wave with P-polarization modulates the superconducting density through the effects of the charge imbalance potential. It is this mechanism of interaction between two polarizations which we plan to exploit in the following consideration of light beams electrodynamics. Note briefly that the above photocurrent density should generate additional Meissner current and superconductor phase distributions at zero frequency. These currents will guarantee the absence of a dc magnetic field deep inside the superconductor and compensate the nonzero divergence of the $z$-component of photocurrent. We will discuss the details of this compensation when considering the electrodynamics of beams below.

\section{\label{sec:four}Interaction of a Bessel beam of twisted light with a superconducting half-space}

As a next step, we develop the description of the interaction of the Bessel beam with the bulk superconductor.
Bessel beam is defined as a superposition of plane waves with wave vectors lying on a cone with a fixed angle to the propagation axis. The magnetic field in the Bessel beam takes the form \cite{Knyasev18}:
\begin{equation}
\label{Besseldef}
\mathbf{H}^{(e)} = B_0 e^{ik_z z}
\int \frac{d^2 \Vec{k}_{\parallel}}{(2\pi)^2} e^{i \Vec{k}_{\parallel} \Vec{\rho}} a_{\kappa m}(k_{\parallel})
\mathbf{e}_\text{H} \ ,
\end{equation}
where 
\begin{equation}
a_{\kappa m}(k_{\parallel}) =
\frac{2\pi}{\kappa} i^{-m} \delta(k_{\parallel} - \kappa) e^{im \varphi_k} \ ,
\end{equation}
$\mathbf{e}_H(\varphi_k) = \alpha \mathbf{e}^\parallel_H(\varphi_k) + \beta \mathbf{e}^\perp_H(\varphi_k)$ is a mixed polarization unity vector, $\varphi_k$ is an angle in the momentum space, and $\Vec{\rho} = (\rho \cos{\varphi}, \rho \sin{\varphi})^T$. 
Details of the calculation of the integral \eqref{Besseldef} are provided in Appendix \ref{Appendix5}.
The incident wave can be written as follows:
\begin{multline}
\label{BesselBeam}
\bH^{(e)}=
\begin{pmatrix}
H_\rho^{(e)} \\
H_\varphi^{(e)} \\
H_z^{(e)}
\end{pmatrix}
=
\frac{B_0}{2} e^{ik_z z + im \varphi} \times
\\
\begin{pmatrix}
-J_{m+1}(\kappa \rho)o_- - J_{m-1}(\kappa \rho) o_+ \\
iJ_{m+1}(\kappa \rho) o_- - iJ_{m-1}(\kappa \rho) o_+ \\
-2\beta \sin\theta J_m(\kappa \rho) 
\end{pmatrix},
\end{multline}
where $ o_\pm = \alpha \pm i\beta \cos\theta$
and $\kappa = k \sin\theta$. The parameter $m$ here can be viewed as a topological quantity which defines the beam's angular momentum. 
Taking into account the reflected wave and continuity of the tangetial component of the magnetic field at the boundary and assuming $r_\parallel,r_\perp \approx 1$, we can neglect the $z$-component of the magnetic field and obtain the field inside the superconductor half-space in the form:
\begin{multline}
\label{BesselBeam}
\bH^{(i)}=
B_0 e^{-z/\lambda_\text{eff} + im \varphi} \times
\\
\begin{pmatrix}
-J_{m+1}(\kappa \rho) o_- - J_{m-1}(\kappa \rho) o_+ \\
iJ_{m+1}(\kappa \rho) o_- - iJ_{m-1}(\kappa \rho) o_+ \\
0
\end{pmatrix} \ .
\end{multline}
Using the boundary conditions at $z=0$ 
\begin{multline}
\partial \tilde{\phi}/\partial z|_{z=0} = -E_z^{(i)}|_{z=0} = -\frac{c}{4\pi\sigma} (\curl \bH^{(i)})_z |_{z=0} = 
\\
-\frac{c}{4\pi\sigma} 2i\alpha \kappa B_0 e^{im \varphi} J_m(\kappa \rho) ,
\end{multline}
\begin{equation}
\partial \Delta_1/\partial z |_{z=0} = 0 
\end{equation}
we solve the equations 
\eqref{phieq}, \eqref{deltaeq} and obtain the expressions for $\tilde{\phi}(\br)$
and $\Delta_1(\br)$:
\begin{equation}
\tilde{\phi} = \frac{i\alpha \omega \sin\theta B_0}{2\pi\sigma_n q_2} e^{-q_2 z + im \varphi}  J_m(\kappa \rho) \ ,
\end{equation}
\begin{multline}
\Delta_1 = -\frac{i\alpha \omega e \tau \nu \Delta_0 B_0\sin\theta}{\pi\hbar\sigma_n \xi^2(q_1^2 - q_2^2)}\times
\\
\left( \frac{e^{-q_1 z}}{q_1} - \frac{e^{-q_2 z}}{q_2} \right) e^{im \varphi} J_m(\kappa \rho).
\end{multline}

In order to obtain the superfluid velocity, we have to calculate $\curl \bH^{(i)}$ and after that use the formula \eqref{vsdef}.  
The main terms of the in-plane components of $\curl\bH^{(i)}$ come from derivatives $\partial/\partial z$, other derivatives are parametrically smaller and  can be neglected:
\begin{multline}
\label{BesselBeam}
\curl\bH^{(i)} \approx
\begin{pmatrix}
-\partial H_\varphi^{(i)}/ \partial z \\
\partial H_\rho^{(i)}/ \partial z \\
\frac{1}{\rho}\partial(\rho H_\varphi^{(i)})/\partial \rho - \frac{1}{\rho}\partial H_\rho^{(i)}/\partial \varphi 
\end{pmatrix}
=
\\
\frac{B_0}{\lambda_\text{eff}} e^{-z/\lambda_\text{eff} + im \varphi}
\begin{pmatrix}
iJ_{m+1}(\kappa \rho) o_- - iJ_{m-1}(\kappa \rho) o_+
\\
J_{m+1}(\kappa \rho) o_- + J_{m-1}(\kappa \rho) o_+ \\
2i\alpha J_m(\kappa \rho) \kappa \lambda_\text{eff}
\end{pmatrix} \ .
\end{multline}

General expression for the photoinduced current of the second order in the incident wave amplitude has two harmonics:
\begin{multline}
\bj_\text{ph} = \bj_\text{ph}^{(0)}  + \bj_\text{ph}^{(2\omega)} = \frac{8e^2aT_c\xi_0^2\Delta_0}{c\hbar^2} \times
\\
\re\left[ \left( \Delta^*_1 + \Delta_1 e^{-2i\omega t}\right) \lambda^2_\text{eff} \curl \bH^{(i)} \right] \ ,
\end{multline}
where $\bj_\text{ph}^{(0)} = -2e\Delta_0\re\left[ \Delta_1^* \mathbf{v}_s \right]$ is the zero-frequency current, 
$\bj_\text{ph}^{(2\omega)} = -2e\Delta_0\re\left[ \Delta_1 \mathbf{v}_s e^{-2i\omega} \right]$ is the second-harmonic response which oscillates at $2\omega$ frequency, and $\mathbf{v}_s$ is given by \eqref{vsdef}.

The resulting photoinduced current at zero frequency takes the form:
\begin{multline}
\label{currentphzero}
\bj^{(0)}_{\text{ph}} = \frac{\nu B_0^2 \omega\sin\theta}{\sqrt{2}H_{cm} \lambda\xi l_E^2}
\re \Bigg[\frac{\lambda^*_\text{eff} f_\text{ph}(z)}{q_1q_2(q_1+q_2)} \times
\\
\begin{pmatrix}
i_1(\rho) + i_2(\rho) 
\\
i_3(\rho) + i_4(\rho) 
\\
i_5(\rho)
\end{pmatrix}
\Bigg] \ ,
\end{multline}
where 
\begin{equation}
f_\text{ph}(z) = \frac{1}{q_1^{-1} - q_2^{-1}} \left( \frac{e^{-q_1 z}}{q_1} - \frac{e^{-q_2 z}}{q_2} \right) \ ,
\end{equation} 
$i_1(\rho) =-|\alpha|^2J_m(\kappa\rho)\frac{\partial J_m(\kappa \rho)}{\kappa\partial\rho}$, 
$i_2(\rho) = \frac{im\alpha\beta^*\cos\theta}{\kappa \rho}J^2_{m}(\kappa \rho)$, 
$i_3(\rho) = \alpha\beta^*\cos\theta J_m(\kappa\rho)\frac{\partial J_m(\kappa \rho)}{\kappa\partial\rho}$, 
$i_4(\rho) = \frac{im|\alpha|^2}{\kappa \rho}J^2_{m}(\kappa \rho)$, $i_5(\rho) = i|\alpha|^2 J_m^2(\kappa \rho) \kappa\lambda^*_\text{eff}$. 

The expression \eqref{currentphzero} contains five contributions to the total current density ($i_{1-5}$): 
\begin{itemize}
    \item The term $i_1$ is proportional to $\frac{\partial J^2_m(\kappa \rho)}{\kappa \partial \rho}$ is a ponderomotive force, since it is determined by the gradient of the intensity of the electromagnetic wave.
    \item The term $i_3$ can be viewed as a manifestation of the inverse Faraday effect and depends on the helicity of the incident wave. The magnetization distribution can be defined as $\mathbf{M}\propto J^2_m(\kappa \rho)$ and the term $J_m(\kappa \rho)\frac{\partial J_m(\kappa \rho)}{\kappa \partial \rho}$ arises, thus, from the current density $c\curl \mathbf{M}$.
    \item Both $i_2$ and $i_4$ terms originate from the photon drag effect. The presence of both radial and azimuthal components indicates the spirality of the drag current. 
    \item The $z$-component $i_5$ is a consequence of the momentum transfer due to the reflection of the incident wave and can be also explained by the photon drag effect.
\end{itemize}

Omitting the calculation details we also give here the expression describing  the second harmonic response: 
\begin{multline}
\label{current2}
\bj^{(2\omega)}_{\text{ph}} = \frac{\nu B_0^2 \omega\sin\theta}{\sqrt{2}H_{cm} \lambda\xi l_E^2}
\re \Bigg[e^{-2i\omega t + 2im\varphi}\frac{\lambda_\text{eff} f_\text{ph}(z)}{q_1q_2(q_1+q_2)} \times
\\
\begin{pmatrix}
i^{(2\omega)}_1(\rho) + i^{(2\omega)}_2(\rho) 
\\
i^{(2\omega)}_3(\rho) + i^{(2\omega)}_4(\rho) 
\\
i^{(2\omega)}_5(\rho)
\end{pmatrix}
\Bigg] \ ,
\end{multline}

where $i^{(2\omega)}_1(\rho) =\alpha^2J_m(\kappa\rho)\frac{\partial J_m(\kappa \rho)}{\kappa\partial\rho}$, 
$i^{(2\omega)}_2(\rho) = \frac{im\alpha\beta\cos\theta}{\kappa \rho}J^2_{m}(\kappa \rho)$, 
$i^{(2\omega)}_3(\rho) = - \alpha\beta\cos\theta J_m(\kappa\rho)\frac{\partial J_m(\kappa \rho)}{\kappa\partial\rho}$, 
$i^{(2\omega)}_4 = \frac{im\alpha^2}{\kappa \rho}J^2_{m}(\kappa \rho)$, $i^{(2\omega)}_5 = \alpha^2 J_m^2(\kappa \rho)\kappa\lambda_\text{eff}$.

The photoinduced dc current calculated above should be obviously screened by the Meissner currents so that to provide the cancellation of the dc magnetic field inside supeconducting halfspace at the distance of the order of the London penetration depth 
$\lambda$ from the surface. Another important point is that dc photocurrents (\ref{currentphzero}) have nozero divergence which must 
be compensated by a certain current which would provide the possibility to avoid the time  -- dependent charge accumulation.
In a non-superconducting systems, this issue is known to be resolved by introducing the inhomogeneous distribution of the chemical potential responsible for the diffusion currents compensating the photoinduced ones \cite{Gunyaga23}. In the case of superconductor an adequate compensating mechanism arises from the inhomogeneous distribution of the superconducting phase.
To calculate all these screening and compensating currents as well as the related magnetic field we consider standard London electrodynamics described by the following equations for the vector potential $\bA^{(0)}$:
\begin{equation}
\label{Meisnersistem}
\begin{cases}
\curl \curl\bA^{(0)} = \frac{4\pi}{c}\left(\bj_\text{M}^{(0)}+\bj_\text{ph}^{(0)}\right),
\\
\frac{4\pi \lambda^2}{c} \bj_\text{M}^{(0)}=- \bA^{(0)} + \frac{\hbar c}{2e} \nabla\chi^{(0)}\ , 
\end{cases}
\end{equation}
where $\curl\bA^{(0)} = \bB^{(0)}$, $\bj^{(0)}_\text{ph}$, $\bj^{(0)}_\text{M}$ and $\chi^{(0)}$ are zero-frequency  quantities. 

From the system \eqref{Meisnersistem} we obtain general equations describing the fields inside and outside the superconductor:
\begin{equation}
\label{Aeqz>0}
\Delta \bA^{(0)} - \grad \dive \bA^{(0)} - \frac{1}{\lambda^2}\bA^{(0)} = -\frac{4\pi}{c}\bj^{(0)}_\text{ph} - \frac{\hbar c}{2e\lambda^2}\nabla \chi^{(0)}
\end{equation}
for $z>0$, and
\begin{equation}
\label{Aeqz<0}
\Delta \bA^{(0)} - \grad \dive \bA^{(0)} = 0
\end{equation}
for $z<0$.

We will solve the problem of calculation of the screening currents in two steps. First, we will calculate the divergence-free component of the current $ j^{(0)}_{\text{M},\varphi} $; this component is essential to determine the vertical component of the magnetic field $ B^{(0)}_z $. Second, we will calculate the component of the current $ j^{(0)}_{\text{M},\rho} $, which obviously has a non-zero divergence. This divergence is compensated  by the vertical component $ j^{(0)}_{\text{M},z} $, driven by the phase gradient $\partial \chi^{(0)}/\partial z$. Using these expressions, we will calculate the in-plain magnetic field component $B^{(0)}_\varphi$.

\textit{The first step: calculation of $j^{(0)}_{\text{M},\varphi}$ and $B_z^{(0)}$.}

\begin{figure}
\centering
\includegraphics[width=0.9\linewidth]{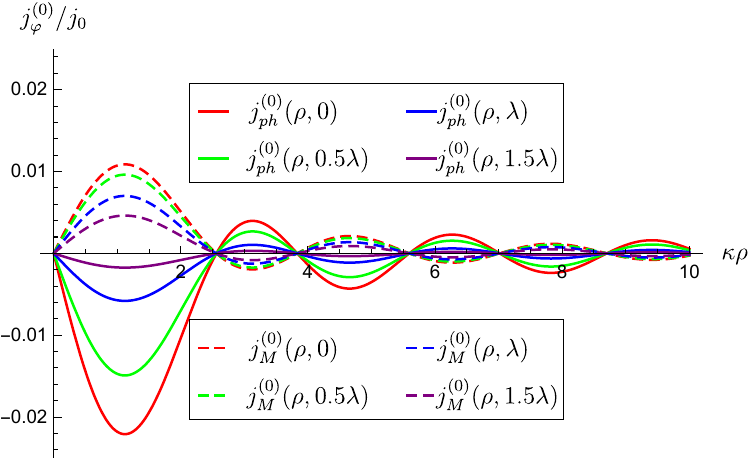}
\caption{Typical photoinduced and Meissner current profiles. Here we put
$\lambda/\xi = 2$, $l_E/\xi = 1/\sqrt{5.79}$, $\omega\tau=1$, $\alpha=1/\sqrt{2}, \beta=i/\sqrt{2}, m=1, \theta=\pi/4$, $j_0=\nu B_0^2 \omega \sin{\theta}/\sqrt{2} H_\text{cm}$.
}
\label{plt4z}
\end{figure}

Let us consider the $\varphi$-component of the photoinduced current  defined by the expression \eqref{currentphzero} :
\begin{equation}
\label{fph}
j^{(0)}_{\text{ph},\varphi}(\rho,z) = \re\left[ f_\text{ph}(z) j^{(c,0)}_{\text{ph},\varphi}(\rho,0) \right] \ ,
\end{equation}
where 
 \begin{equation}
\label{jc0}
j^{(c,0)}_{\text{ph},\varphi}(\rho,0) = \frac{\nu B_0^2\omega\sin\theta}{\sqrt{2}H_\text{cm}l_E^2\xi\lambda} \frac{\lambda^*_\text{eff}}{q_1 q_2(q_1+q_2)}(i_3(\rho)+i_4(\rho)) \ .
\end{equation}
Note that due to the cylindrical symmetry the equation \eqref{Aeqz>0} for the $\varphi$-component of $\bA$ can be written as:
\begin{equation}
\label{Avarphieqsimplified}
\begin{cases}
(\Delta \bA^{(0)})_\varphi - \frac{1}{\lambda^2}A^{(0)}_\varphi = -\frac{4\pi}{c}j^{(0)}_{\text{ph},\varphi}(\rho,z) \ , z>0 \ ,
\\
(\Delta \bA^{(0)})_\varphi = 0 \ , z<0 \ .
\end{cases}
\end{equation}
Accounting for the difference between the length scales of the solution in the plane $(xy)$ and along the $z$ direction ($1/\kappa \gg \xi, \lambda$) we can search the vector potential in the form:
\begin{equation}
\label{Avarphi}
A^{(0)}_\varphi = -\frac{4\pi\lambda^2}{c} \re \left[ f_\text{M}(z) j^{(c,0)}_{\text{ph},\varphi}(\rho,0) \right] \ .
\end{equation}
Taking now the equations
\begin{equation}
\label{Avarphieqsimplified2}
\begin{cases}
\frac{\partial^2 A^{(0)}_\varphi}{\partial z^2} - \frac{1}{\lambda^2}A^{(0)}_\varphi = -\frac{4\pi}{c}j^{(0)}_{\text{ph},\varphi}(\rho,z) \ , z>0 \ ,
\\
\frac{\partial^2 A^{(0)}_\varphi}{\partial z^2} = 0 \ , z<0 \ ,
\end{cases}
\end{equation}
and assuming the continuity of the function $A^{(0)}_\varphi$ and its derivative at $z = 0$ we find the solution which does not grow at $z = \pm \infty $:
\begin{multline}
\label{Avarphi2}
f_\text{M}(z) = \frac{\lambda^{-2}}{q_2 - q_1}\left( \frac{q_2 e^{-q_1 z}}{q_1^2 - \lambda^{-2}} - \frac{q_1 e^{-q_2 z}}{q_2^2 - \lambda^{-2}} \right) -
\\
-\frac{\lambda^{-1}}{q_1^{-1}-q_2^{-1}} \left(\frac{1}{q_1^2 - \lambda^{-2}} - \frac{1}{q_2^2 - \lambda^{-2}} \right) e^{-z/\lambda} \ .
\end{multline}
The spatial profiles of the resulting photoinduced and Meissner currents are shown in Fig. \ref{plt4z}.

It is useful to note here that due to simple relation
\begin{equation}
\int_0^{+\infty}dz f_\text{ph}(z) = - \int_0^{+\infty}dz f_\text{M}(z) = \frac{q_1+q_2}{q_1q_2}.
\end{equation}
 the  resulting magnetic moment of the total current is equal to zero:
\begin{equation}
M_z = \frac{1}{c}\int_0^\infty\int_0^\infty dz d\rho \, \pi \rho^2 \left(j_{\text{ph},\varphi}^{(0)} + j_{\text{M},\varphi}^{(0)}\right) = 0 \ .
\end{equation}
Thus, magnetic moments of the photoinduced and Meissner current exactly compensate each other.

\begin{figure*}[t]
\centering
\includegraphics[width=0.9\linewidth]{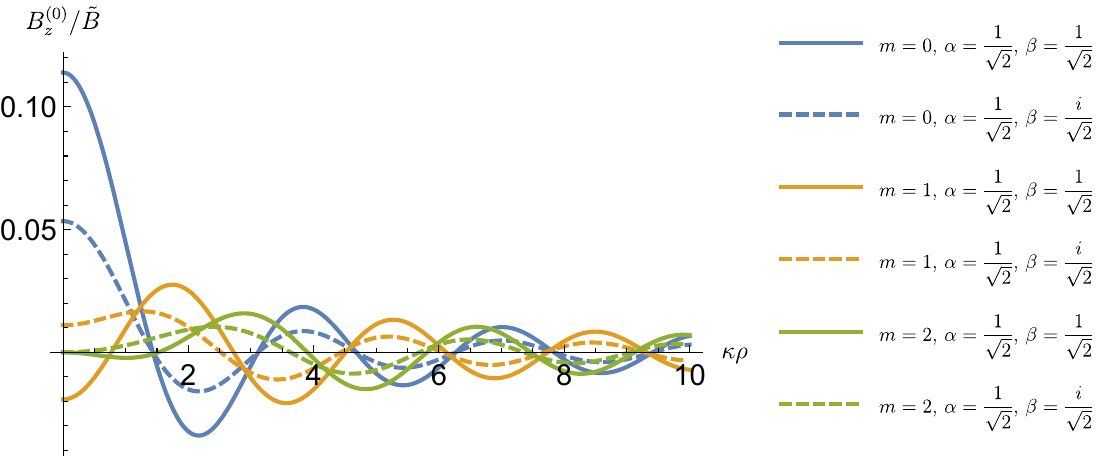}
\caption{Magnetic field profiles $B_z(\rho)$. Here we put $\lambda/\xi = 2$, $l_E/\xi = 1/\sqrt{5.79}$, $\omega\tau=1$,  $\tilde{B}=4\pi\nu B_0^2/\sqrt{2} H_\text{cm}$. The values $\alpha, \beta, m$ are shown in the legend.}
\label{Bz012}
\end{figure*}

\begin{figure*}[t]
\centering
\includegraphics[width=0.9\linewidth]{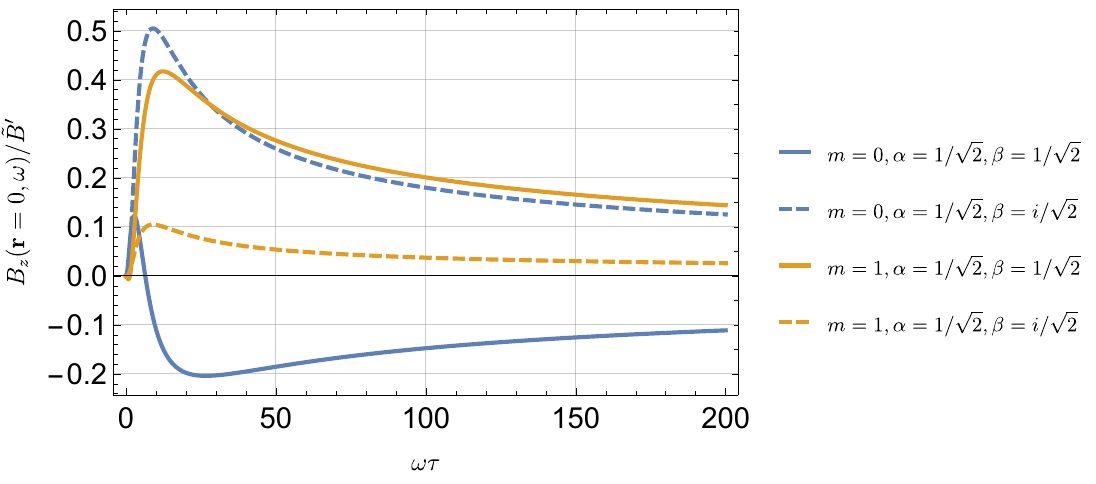}
\caption{Frequency dependence of magnetic field $B_z(\omega)$. Here we put $\lambda/\xi = 2$, $l_E/\xi = 1/\sqrt{5.79}$, $\omega\tau=1$,  $\tilde{B}'=4\pi\nu (\lambda/c\tau)^2 B_0^2/\sqrt{2} H_\text{cm}$. The values $\alpha, \beta, m$ are shown in the legend. At $\omega\tau \to \infty$ the magnetic field decays as $\propto 1/\sqrt{\omega\tau}$.
}
\label{pltomegatau}
\end{figure*}

Keeping in mind possible experiments on photon drag effects it looks reasonable to focus on the calculation of magnetic field components which can be measured by different local probe or magnetooptics techniques \cite{Anahory2014, Hovhannisyan2025, Hicks2007, MOIbook}.
We start from the analysis
of the profiles of the $B_z$ component of the magnetic field determined by the $\varphi$ component of the vector potential
\begin{equation}
\label{Bz}
B^{(0)}_z = \frac{1}{\rho}\frac{\partial (\rho A^{(0)}_\varphi)}{\partial \rho} = -\frac{4\pi\lambda^2}{c}f_\text{M}(z) \frac{\partial(\rho j^{(0)}_{\text{ph},\varphi}(\rho,0))}{\rho \partial \rho} \ .
\end{equation} 
 The typical spatial distributions of the $z$-component of the magnetic field are shown in Fig.\ref{Bz012}. 
Note that within the above approximation  $1/\kappa \gg \xi, \lambda$ the value $B^{(0)}_z(0,0)$ appears to be nonzero  only for $m=0,\pm 1$.
The frequency dependence of the magnetic field in the center of the beam is shown in Fig.\ref{pltomegatau}. This plot can be also viewed as a temperature dependence of the field since the frequency enters the related expressions through the factor $\omega\tau$, where $\tau \sim 1/(T_c-T)$. Experimentally temperature dependences are probably more accessible for measurements than the frequency ones as the latter assume the tuning of the frequency of the incident wave. 


In order to obtain the nonzero value of $B^{(0)}_z(0,0)$ for other $m$ numbers, we need
to consider the magnetic field up to  the first order corrections in the parameter $\kappa\lambda$.
Using the Fourier-Bessel transform
\begin{align}
A^{(0)}_\varphi(k_\rho) &= \int_0^{+\infty} A^{(0)}_\varphi(\rho)J_1(k_\rho \rho)\rho d\rho \ ,
\\
j_{\text{ph},\varphi}^{(0)}(k_\rho) &= \int_0^{+\infty} j_{\text{ph},\varphi}^{(0)}(\rho)J_1(k_\rho \rho)\rho d\rho \ .
\end{align}
in Eqs. \eqref{Avarphieqsimplified}, \eqref{Avarphieqsimplified2} we get the equations:
\begin{equation}
\begin{cases}
\frac{\partial^2 A^{(0)}_\varphi}{\partial z^2} - \left(k_\rho^2 + \frac{1}{\lambda^2}\right)A^{(0)}_\varphi = -\frac{4\pi}{c}f_\text{ph}(z)j^{(0)}_{\text{ph},\varphi}(k_\rho,0) \ , z>0 \ ,
\\
\frac{\partial^2 A^{(0)}_\varphi}{\partial z^2} - k_\rho^2 A^{(0)}_\varphi = 0 \ , z<0 \ .
\end{cases}
\end{equation}

In the first order of the perturbation theory in the parameter $\kappa\lambda$ we obtain a nonzero magnetic field in the center for $|m|>1$:
\begin{multline}
\label{Bcenter}
B^{(0)}_z(0,0) = -\frac{2\sqrt{2}\pi\nu B_0^2 \kappa^3 \lambda^2}{H_{cm} \xi l_E^2} \times 
\\
\re \left[\frac{\lambda^*_\text{eff}(1 + (q_1 + q_2)\lambda)}{q_1 q_2 (q_1 + q_2)(1 + q_1 \lambda)(1 + q_2 \lambda)} C \right] \approx
\\
-2\sqrt{2}\pi\frac{\nu B_0^2 \kappa^3 \lambda^2}{H_{cm} \xi l_E^2} \re \Biggl[ \frac{\lambda^*_\text{eff}}{q_1^2 q_2^2} C \Biggr] \ ,
\end{multline}
where 
\begin{multline}
\label{Cdef}
C = \int_0^{+\infty} \int_0^{+\infty}  \rho d \rho \frac{k_\rho^3}{\kappa^2} d k_\rho J_1(k_\rho \rho) (i_3(\rho)+i_4(\rho)) =
\\
= \frac{32(im|\alpha|^2 -(3\pi/2)\alpha\beta^*\cos\theta) }{\pi(4m^2-1)(4m^2-9)} \ .
\end{multline}
is a complex number of the order unity.

Note that considering the generalization of the above perturbation theory for the full $B^{(0)}_z(\rho,0)$ profiles (not only for $\rho=0$) we need to take account of small corrections to the reflection coefficients. 

\textit{The second step: calculation of $j^{(0)}_{\text{M},\rho}, j^{(0)}_{\text{M},z}$ and $B^{(0)}_\varphi$.}

Let us now consider the vector-potential component induced by the radial and vertical ($z$) components of the photocurrent and take the expansion of this quantity in the small parameter $\kappa\lambda$: $\bA^{(0)}_\text{div} = \bA^{(0)}_0+\mathbf{a}^{(0)}$, where $\bA^{(0)}_0 = (A^{(0)}_0,0,0)^T$ being the main term and $\mathbf{a}^{(0)}$ being the higher-order correction.
In the leading approximation in the parameter $\kappa\lambda$ the corresponding equations take the form
\begin{equation}
\label{A0}
\begin{cases}
\frac{\partial^2 A^{(0)}_0}{\partial z^2} - \frac{1}{\lambda^2}A^{(0)}_0 = -\frac{4\pi}{c}j^{(0)}_{\text{ph},\rho}(\rho,z) \ , z>0 \ ,
\\
\frac{\partial^2 A^{(0)}_0}{\partial z^2} = 0 \ .
\end{cases}
\end{equation} 

The solution of Eq. \eqref{A0} is similar to Eq. \eqref{Avarphi}:
\begin{equation}
\label{A0sol}
A^{(0)}_0 = -\frac{4\pi\lambda^2}{c} \re \left[ f_\text{M}(z) j^{(c,0)}_{\text{ph},\rho}(\rho,0) \right] \ .
\end{equation}

The total radial current can be written as follows 
\begin{equation}
j^{(0)}_{\text{sum},\rho} = 
\re \Bigl[ (f_\text{ph}(z)+f_\text{M}(z))j^{(c,0)}_{\text{ph},\rho}  \Bigr] \label{jsumrho}
\end{equation}
while the $j^{(0)}_{\text{sum},z}$ current component can be obtained by solving the equation $\dive \bj^{(0)}_\text{sum} = \frac{\partial (\rho j^{(0)}_{\text{sum},\rho})}{\rho \partial \rho} + \frac{\partial j^{(0)}_{\text{sum},z}}{\partial z} = 0$. Integrating over $z$ we find:
\begin{equation}
j^{(0)}_{\text{sum},z} = - 
\re \Bigg[\left(\int_0^z (f_\text{ph}(z)+f_\text{M}(z))dz \right) \frac{\partial (\rho j^{(c,0)}_{\text{sum},\rho}(\rho,0))}{\rho \partial \rho} \Bigg] \ ,
\end{equation}
with 
\begin{multline}
\int_0^z (f_\text{ph}(z)+f_\text{M}(z))dz = 
\\
-\frac{1}{q_1^{-1}-q_2^{-1}}\left( \frac{e^{-q_1 z} - e^{-z/\lambda}}{q_1^2 - \lambda^{-2}} - \frac{e^{-q_2 z} - e^{-z/\lambda}}{q_2^2 - \lambda^{-2}} \right) \ .
\end{multline}

The azimuthal magnetic field can be found by the formula 
\begin{equation}
\label{Bphi0}
B^{(0)}_\varphi \approx \frac{\partial A^{(0)}_0}{\partial z} \ .
\end{equation}

\begin{figure}
\centerline{\includegraphics[width=1.0\columnwidth]{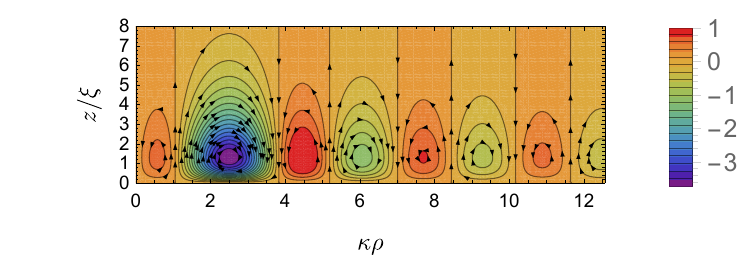}}
\caption{The colorplot of the magnetic field component $B^{(0)}_\varphi(\rho,z)/\tilde{B}$ and the corresponding current lines. Here we put $m=1$, $\alpha=1/\sqrt{2}$, $\beta = i/\sqrt{2}$,  $\xi/\lambda=1/2$, $\xi/l_E = \sqrt{5.79}$, $\omega\tau=1$, $\kappa\lambda=2\cdot10^{-3}$, $\theta=\pi/4$. The isolines of the $B^{(0)}_\varphi$ profiles coincide with the current lines $\bj^{(0)}_\text{sum}$.}
\label{BphiPlot}
\end{figure}

The resulting colorplot of the magnetic field $B^{(0)}_\varphi$ and the corresponding current lines are shown in Fig. \ref{BphiPlot}. Thus, the total current distribution has a toroidal shape, i.e., the currents flow both in the azimuthal direction and along closed paths in the $(\rho, z)$ plane. It is also interesting to note that due to the strong current anisotropy the magnetic field $B^{(0)}_\varphi$--component strongly exceeds the  $B^{(0)}_z$--component in the large parameter $(\kappa\lambda)^{-1}$.

Let us finally briefly comment on the behavior of the small correction to the vector potential $\mathbf{a}^{(0)}$ and the phase gradient needed to get zero divergence of the total current. 
The equation for $\mathbf{a}^{(0)}$ reads:
\begin{equation}
\label{aeq}
\Delta \mathbf{a}^{(0)} - \frac{1}{\lambda^2}\mathbf{a}^{(0)} = \frac{\partial}{\partial z}\left( \frac{1}{\rho}\frac{\partial (\rho A^{(0)}_0)}{\partial \rho}\right)\mathbf{e}_z - \nabla \chi^{(0)} - \frac{4\pi}{c} j^{(0)}_{\text{ph},z} \mathbf{e}_z
\end{equation}
Here we impose the gauge $\dive \mathbf{a}^{(0)} = 0$ and $a_z=0$. Taking the divergence of the left- and right-hand sides of the equation \eqref{aeq} we get the Laplace equation by $\chi^{(0)}$:
\begin{equation}
\Delta \chi^{(0)} = - \frac{4\pi}{c}\frac{\partial j^{(0)}_{\text{ph},z}}{\partial z} + \frac{\partial^2}{\partial z^2}\left( \frac{1}{\rho} \frac{\partial (\rho A^{(0)}_0)}{\partial \rho} \right)
\end{equation}
The main term of the phase $\chi^{(0)}$ can be found integrating the above equation  over $z$ twice (the constant of the integration $\chi^{(0)}(\infty)=0$):
\begin{multline}
\label{chi}
\chi^{(0)}(\rho,z) = -\frac{4\pi}{c}\re \Bigg[ \int_{+\infty}^z f_\text{ph}(z')dz' \cdot j^{(c,0)}_{\text{ph},z}(\rho,0) + 
\\
+ f_\text{M}(z)\lambda^2\frac{\partial (\rho j^{(c,0)}_{\text{ph},\rho}(\rho,0))}{\rho\partial \rho}
\Bigg]
\end{multline}
As noted above all these corrections are generally important to guarantee the absence of the current sources but do not provide essential contributions to the light induced magnetic fields.

\section{\label{sec:five}Interaction of a Bessel beam of twisted light with a superconducting thin film}

Both linear and nonlinear electrodynamics considered above should be strongly affected by the sample geometry and, in particular, by its thickness. As soon as the thickness becomes comparable with the characteristic length scales one can expect the modification of current spatial distributions and resulting magnetic fields. 
Thus, keeping in mind possible experimental setup it 
is reasonable to generalize the above results for the case of a thin superconducting film of the thickness $d$  less than the London penetration depth.
Similarly to the above analysis we need first to describe the generation of photocurrents by the incident plain wave.
The expressions for the reflection coefficient similar to Eqs.\eqref{r_parallel}, \eqref{r_perp} for P- and S-polarizations take the form (see \cite{LL8}):
\begin{align}
r_\parallel &= \frac{1}{1 - \frac{2i \omega \lambda^2_\text{eff}}{c \cos{\theta}d}}, 
\\
r_\perp &= \frac{1}{1 - \frac{2i\cos{\theta} \omega \lambda^2_\text{eff}}{c d} }. 
\end{align}
Hereafter we assume that the film is not too thin, so that $\omega\lambda^2/c d \ll 1$ and, thus, the reflection coefficient appears to be close to unity and the magnetic field at the second boundary of the film is equal to zero. Taking $\omega \sim  10^{12} s^{-1}$ and $\lambda \sim 100\ nm$, we find that the above condition breaks down only when $d \lesssim 1\ nm$. Thus, we assume the film to be thicker than the atomic length scale.

A significant difference between the film and the bulk superconductor is the presence of a second boundary, which implies additional boundary conditions for the functions $H(z)$, $\tilde{\phi}(z)$ and $\Delta_1(z)$.  Starting again from the case of the incident plane wave we find the resulting magnetic field distribution described by the expressions: 
\begin{equation}
\bH(z) = \mathbf{e}_{\text{H}} B_0 e^{ik_x x} \times
\begin{cases}
\left( e^{ik_z z} + e^{-ik_z z} \right) , z<0 \ ,
\\
2\frac{\sinh{\frac{d-z}{\lambda_\text{eff}}}}{\sinh{\frac{d}{\lambda_\text{eff}}}} , 0<z<d \ ,
\\
0 , z>d \ ,
\end{cases}
\end{equation}
where $\mathbf{e}_{\text{H}} = (\beta\cos\theta, \alpha,0)^T$. By analogy to the above calculation in the bulk we obtain both $\tilde{\phi}(z)$ (with two boundary conditions $\partial \tilde\phi /\partial z|_{z=0} = -E_z(z=0)=-\frac{i\omega\sin\theta}{2\pi\sigma_n}B_0$, $\partial \tilde\phi /\partial z|_{z=d}=-E_z(z=d)=0$) 
and $\Delta_1(z)$ (with two boundary conditions $\partial \Delta_1/\partial z = 0$ at $z=0,d$):
\begin{equation}
\tilde{\phi}(z) = \alpha e^{i k_x x} \frac{2i \omega \sin\theta B_0}{4\pi\sigma_n q_1} \frac{\cosh{q_1(d-z)}}{\sinh{q_1 d}}
\end{equation}
\begin{multline}
\Delta_1(z) = \alpha e^{i k_x x} \frac{4e\tau\nu\Delta_0}{\hbar \xi^2} \frac{i \omega \sin\theta B_0} {4\pi\sigma_n (q_2^2 - q_1^2)} 
\times \\ 
\left( \frac{\cosh{q_1(d-z)}}{ q_1\sinh{q_1 d}} - \frac{\cosh{q_2(d-z)}}{q_2\sinh{q_2 d}} \right)
\end{multline}

Taking into account the $\Delta_1(z)$ dependence  after integration over $z$ from 0 to $d$ we get the total current:
\begin{multline}
\mathbf{I}^\text{film}_{\text{ph},\parallel} = \frac{\lambda}{d}\frac{\nu}{\sqrt{2}\pi} \frac{\xi}{\tau} \frac{B_0^2\sin{\theta}}{H_{cm}} \frac{\omega \tau}{1+\omega^2\tau^2\eta^2} \times
\\
\re\left[ \frac{i}{\sqrt{1-i\omega\tau\eta}}\frac{\sqrt{1+i\omega\tau\eta}}{2-i\omega\tau} \begin{pmatrix}
|\alpha|^2 \\ -\alpha \beta^* \cos{\theta} \\ -i|\alpha|^2\sin\theta \frac{\lambda\omega}{c\sqrt{1+i\omega\tau\eta}}
\end{pmatrix} \right].
\end{multline}

Further procedure of calculation of the photocurrent at zero frequency in the film for the case of the Bessel beam is straightforward:
\begin{multline}
\label{current0film}
\bj^{(f,0)}_\text{ph} = \frac{\lambda}{d}\frac{\nu B_0^2 \omega\sin\theta}{\sqrt{2}H_{cm}\xi \lambda^2 l_E^2} 
\re \Bigg[\frac{\lambda^2_\text{eff}}{q_1^2-q_2^2} \times
\\
\left((q_1 \tanh{q_1 d})^{-1} - (q_2 \tanh{q_2 d})^{-1}\right)f^{(f)}_\text{ph}(z) \times
\\
\begin{pmatrix}
i_1(\rho) + i_2(\rho) 
\\
i_3(\rho) + i_4(\rho)
\\
i_5(\rho)
\end{pmatrix}
\Bigg]
\end{multline} 
where $i_{1-5}(\rho)$ are given above (after the expression \eqref{currentphzero}), and

\begin{multline}
f^{(f)}_\text{ph}(z) = \frac{1}{(q_1 \tanh{q_1 d})^{-1} - (q_2 \tanh{q_2 d})^{-1}} \times
\\
\Biggl(\frac{\cosh q_1(d-z)}{q_1\sinh q_1 d}  
-\frac{\cosh q_2(d-z)}{q_2\sinh q_2 d}\Biggr) \ .
\end{multline}

\begin{figure}
\centering
\includegraphics[width=0.9\linewidth]{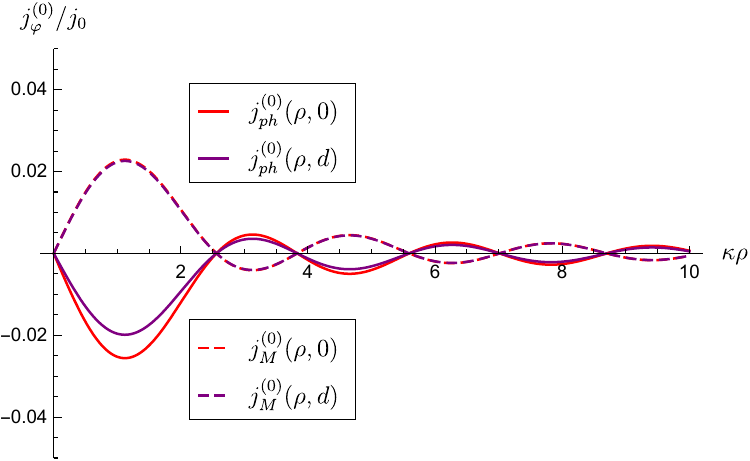}
\caption{Typical photoinduced and Meissner current profiles for the film. Here we put $d=3\lambda/4$,
$\lambda/\xi = 2$, $l_E/\xi = 1/\sqrt{5.79}$, $\omega\tau=1$,  $\alpha=1/\sqrt{2}, \beta=i/\sqrt{2}, m=1, \theta=\pi/4$, $j_0=\nu B_0^2 \omega \sin{\theta}/\sqrt{2} H_\text{cm}$.}
\label{plt4zFilm}
\end{figure}

\begin{figure}
\centerline{\includegraphics[width=1.0\columnwidth]{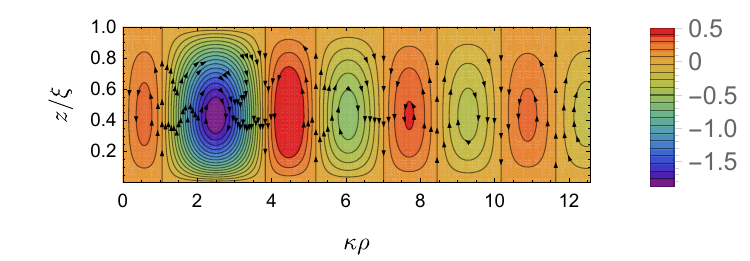}}
\caption{The colorplot of the magnetic field component $B^{(0)}_\varphi(\rho,z)/\tilde{B}$ and the corresponding current lines. Here we put $m=1$, $\alpha=1/\sqrt{2}$, $\beta = i/\sqrt{2}$,  $d/\lambda=1/2$, $\xi/\lambda=1/2$, $\xi/l_E = \sqrt{5.79}$, $\omega\tau=1$, $\kappa\lambda=2\cdot10^{-3}$, $\theta=\pi/4$. The isolines of the $B^{(0)}_\varphi$ profiles coincide with the current lines $\bj^{(f,0)}_\text{sum}$.}
\label{CurrentsFilm}
\end{figure}

The components of the Meissner screening current are described by the expressions:
\begin{equation}
\bj^{(f,0)}_{\text{M}}(\rho,z) = -\frac{c}{4\pi\lambda^2}\bA^{(f,0)} = \re \Bigg[ f^{(f)}_\text{M}(z) \bj^{(c,f,0)}_{\text{ph}}(\rho,0) \Bigg]
\end{equation}
with 
\begin{multline}
f^{(f)}_\text{M}(z) = \frac{\lambda^{-2}}{(q_1 \tanh{q_1 d})^{-1} - (q_2 \tanh{q_2 d})^{-1}} \times
\\
\Biggl(\frac{\cosh q_1(d-z)}{q_1Q_1\sinh q_1 d}  
-\frac{\cosh q_2(d-z)}{q_2Q_2\sinh q_2 d} -
\\
\frac{\lambda^2}{d} \left(\frac{1}{Q_1}-\frac{1}{Q_2} \right) \Biggr)
\end{multline}
and $Q_{1,2} = q_{1,2}^2-\lambda^{-2}$.
The spatial profiles of the resulting photoinduced and Meissner currents in a thin film are shown in Fig. \ref{plt4zFilm}.
Note that for a thin film with $d \ll \lambda$ the current density is almost homogeneous along the $z$-coordinate.
Using the formula \eqref{Bphi0}, we calculate also the magnetic field $\bB=\curl \bA$. 
$B^{(0)}_\varphi(\rho,z)$ distribution in the film and the current flow are shown in Fig. \ref{CurrentsFilm}.

Note that the $B^{(0)}_z(\rho)$ profiles coincide with those for superconducting half-space up to a certain constant depending on the film thickness. The resulting vertical magnetic field in the film is of the order:
\begin{equation}
B^{(0)}_z \sim \nu \frac{\lambda^2}{d^2} \kappa^2 \lambda\xi \frac{B_0^2}{H_\text{cm}}
\end{equation}
which is larger than the half-space result by the factor $(\lambda/d)^2$ which can also be written as a ratio of the Pearl length $\lambda_\text{Pearl}=\lambda^2/d$ to the film thickness. The Pearl length is well known as a natural length scale determining the screening properties of thin superconducting films \cite{Pearl64}.
The first $\lambda/d$ multiplier in this product results from the increase of the superfluid velocity $\mathbf{v}_s$ at frequency $\omega$; the second $\lambda/d$ multiplier originates from the boundary conditions for the dc vector potential (following from the absence of the dc magnetic field outside the superconductor). The total factor $(\lambda/d)^2$  can strengthen the resulting photoinduced magnetic fields and, thus, facilitate the experimental observation of the above effects.

\section{\label{conclusion} Conclusion} 

To sum up, we have developed a phenomenological theory describing the interaction of structured electromagnetic radiation, specifically Bessel beams of twisted light, with superconducting systems and demonstrated that such beams can induce dissipationless dc photocurrents and associated magnetic fields that strongly depend on both the helicity and orbital angular momentum of light.

Our analysis revealed two mechanisms of the photoinduced magnetic response: (i) the transfer of the orbital momentum of the electromagnetic wave to the Cooper pairs; 
(ii) the inverse Faraday effect  for the superconducting condensate caused by the helicity of the beam. The first contribution to the photocurrent is an odd function of the angular momentum quantum number 
$m$, while the second one is even in 
$m$. 
Both terms arise from the second order nonlinear response of the Cooper pairs resulting from the generation of the nonequilibrium charge imbalance potential which in turn affects the superfluid density due to a small particle-hole asymmetry.

We also provided detailed estimates of screening Meissner currents and derived explicit spatial profiles of induced magnetic fields, offering practical guidance for experimental detection using various magnetometry techniques. 
The presented results lay the groundwork for the experimental exploration of light-induced persistent currents in superconductors using structured light. 

\begin{acknowledgements}
The authors thank S. V. Mironov, A. I. Buzdin and M. V. Durnev for stimulating and useful duscussions. This work was supported by the Russian Science Foundation (Grant No. 25-12-00042) in part of the analysis of the case of half-space and by the Grant of the Ministry of science and higher education of the Russian Federation No. 075-15-2025-010 in part of calculations for the case of a thin film. 

\end{acknowledgements}

\appendix

\section{\label{Appendix1} Derivation of basic equations of perturbation theory in TDGL model}

The aim of this section is to derive the equations for $\tilde{\phi}(\br)$ and $\Delta_1(\br)$ starting from basic TDGL equation \eqref{GL-equationAp}:
\begin{multline}
\label{GL-equationAp}
-\Gamma\left(\hbar \frac{\partial}{\partial t} - 2ie\phi\right)\Delta =
\\
- a T_c \epsilon\Delta + b |\Delta|^2\Delta + a T_c \xi_0^2 \left(-i \mathbf{\nabla} + \frac{2\pi}{\Phi_0} \mathbf{A} \right)^2\Delta.
\end{multline}

Let us introduce a new variable $ \tilde{\phi} = \phi - \frac{\hbar}{2e} \partial\chi/\partial t  $.
Taking the real part of Eq. \eqref{GL-equationAp} we get:
\begin{equation}
-\tau \frac{\partial\Delta_1}{\partial t} + \frac{2e}{\hbar} \tau \nu \Delta_0 \tilde{\phi} = 2\Delta_1 - \xi^2 \nabla^2 \Delta_1.
\end{equation}

After replacing $\partial/\partial t \rightarrow -i\omega$ we get:
\begin{equation}
\label{DeltaeqAp}
-\nabla^2 \Delta_1 + q_1^2 \Delta_1 = \frac{2e\tau \nu \Delta_0}{\hbar \xi^2} \tilde{\phi},
\end{equation}
where
$ q_1^2 \equiv \frac{2-i\omega\tau}{\xi^2}$.

Multiplying the imaginary part of the Ginzburg-Landau equation \eqref{GL-equationAp} by the factor $4e\Delta^*/\hbar$ we find: 
\begin{multline}
\label{current1}
\dive \mathbf{j}_s = \Delta_0^2 \frac{4a T_c \xi_0^2 \Delta_0^2 e}{\hbar}\left(\Delta\chi - \frac{2\pi}{\Phi_0} \dive\mathbf{A} \right) = 
\\
\frac{\pi a \Delta_0^2 e^2}{\hbar} \tilde{\phi}.
\end{multline}

The continuity equation reads:

\begin{equation}
\label{current2}
\dive \mathbf{j}_s = -\dive \mathbf{j}_n = \sigma_n \dive \left(\mathbf{\nabla}\phi + \frac{1}{c} \partial \mathbf{A}/\partial t\right).
\end{equation}

Adding and subtracting $\pm \frac{\hbar}{2e}\nabla(\partial\chi/\partial t)$ in \eqref{current2} and taking into account \eqref{current1} and \eqref{current2} together, we obtain:
\begin{equation}
\Delta\tilde{\phi} - \frac{\pi \hbar} {8 T_c \xi_0^2} \partial\tilde{\phi}/\partial t = \frac{\pi a \Delta_0^2 e^2} {\hbar \sigma_n} \tilde{\phi}.   
\end{equation}

Substituting $\partial/\partial t \rightarrow -i\omega$ we can write the final equation for the charge imbalance potential:
\begin{equation}
\label{phieqAp}
\Delta \tilde{\phi} = q_2^2 \tilde{\phi},
\end{equation}
where
$ q_2^2 \equiv \frac{1}{l_E^2} - \frac{i\omega\tau}{\xi^2} = \frac{\lambda^2}{l_E^2 \lambda_\text{eff}^2}$.

\section{\label{Appendix2} Basic equations for electric and magnetic fields}

Maxwell's equations for the electromagnetic field inside the superconductor read:
\begin{gather}
\label{Maxwell1}
\curl\mathbf{H}_i = \frac{4\pi\sigma_n}{c} \mathbf{E}_i + \frac{4\pi}{c} \mathbf{j}_s + \frac{1}{c} \frac{\partial \mathbf{D}_i}{\partial t}
\\
\label{Maxwell2}
\curl\mathbf{E}_i = -\frac{1}{c} \frac{\partial \mathbf{H}_i}{\partial t}
\end{gather}
where $\bD_i = \varepsilon \bE_i$.

Using the London theory and exploiting the relation $\frac{\partial}{\partial t}\left(A - \frac{\Phi_0}{2\pi}\nabla \chi \right) \rightarrow -\frac{ic}{\omega} \left(\bE + \nabla \tilde{\phi}\right) $ we find the expression for the current density written through the complex amplitudes:
\begin{equation}
\mathbf{j}_s = -\frac{c}{4\pi \lambda^2} \left(\bA - \frac{\Phi_0}{2\pi}\nabla \chi\right) = i \frac{c^2}{4\pi\omega \lambda^2 } \left(\mathbf{E}_i + \nabla \tilde{\phi}\right) \ .
\end{equation}

After the Fourier transform in time variable we obtain the equation for the magnetic field:
\begin{equation}
\label{Heq}
\Delta \mathbf{H}_i = \left(\frac{1}{\lambda^2} - \frac{4\pi i\sigma_n \omega}{c^2} - \frac{\omega^2}{c^2}\varepsilon\right) \mathbf{H}_i
\end{equation}

Neglecting the last term in parentheses and introducing the effective penetration depth
\begin{equation}
\frac{1}{\lambda_{\text{eff}}^2} = \frac{1}{\lambda^2} - \frac{4\pi i\sigma_n \omega}{c^2} = \frac{1}{\lambda^2} \left(1-i\omega\tau \frac{l_E^2}{\xi^2}\right),
\end{equation}
we get the equations used in the main text:
\begin{equation}
\label{Heqshort}
\Delta \mathbf{H}_i = \frac{1}{\lambda_{\text{eff}}^2} \mathbf{H}_i \ ,
\end{equation}

\begin{equation}
\label{Eeq}
\curl\mathbf{H} = \frac{ic}{\omega} \left(\frac{1}{\lambda_\text{eff}^2} \bE + \frac{1}{\lambda^2} \nabla\tilde{\phi}\right) \ .
\end{equation}

\section{\label{Appendix5}Field of twisted light in vacuum}

In this section, we calculate the magnetic field of the incident twisted light in vacuum. The incident wave has two polarizations, so we introduce the polarization unit vector corresponding to the magnetic field direction:
\begin{equation}
\mathbf{e}_\text{H} = \alpha \mathbf{e}_\text{H}^{\parallel} + \beta \mathbf{e}_\text{H}^{\perp},
\end{equation}
where $ |\alpha|^2 + |\beta|^2 = 1 $, and $ \sin\theta = \frac{\kappa}{k} $.

\begin{equation}
\mathbf{k}/k = (\sin\theta \cos\varphi_k, \sin\theta \sin \varphi_k, \cos\theta)^T,
\end{equation}
where $k^2 = k_\parallel^2 + k_z^2$, $\varphi_k \in [0,2\pi)$.
Unit vectors for P-and S-polarizations can be written as follows:
\begin{equation}
\begin{cases}
\mathbf{e}_\text{H}^{\parallel} = (-\sin\varphi_k, \cos\varphi_k, 0)^T,
\\
\mathbf{e}_\text{H}^{\perp} = (\cos\theta \cos\varphi_k, \cos\theta \sin\varphi_k, -\sin\theta)^T.
\end{cases}
\end{equation}
Therefore, the complex amplitudes of the magnetic field of the incident wave $\mathbf{H}^{(e)} $ take the form:
\begin{equation}
\mathbf{H}^{(e)} = B_0 e^{ik_z z}
\int \frac{d^2 \Vec{k}_{\parallel}}{(2\pi)^2} e^{i \Vec{k}_{\parallel} \Vec{\rho}} a_{\kappa m}(k_{\parallel})
\mathbf{e}_\text{H}.
\end{equation}
where
\begin{equation}
a_{\kappa m}(k_{\parallel}) =
\frac{2\pi}{\kappa} i^{-m} \delta(\kappa - k_{\parallel}) e^{im \varphi_k}.
\end{equation}

In Cartesian coordinates the expression for  the magnetic field can be rewritten as follows:
\begin{widetext}
\begin{multline}
\bH^{(e)} =
\begin{pmatrix}
H_x^{(e)} \\
H_y^{(e)} \\
H_z^{(e)}
\end{pmatrix} =
\frac{B_0}{2\pi}
i^{-m} e^{ik_z z}
\int d \varphi_k  e^{i \mathbf{k}_{\parallel} \Vec{\rho}} e^{im \varphi_k}
\begin{pmatrix}
-\alpha \sin \varphi_k + \beta \cos \theta \cos \varphi_k
\\
\alpha \cos \varphi_k + \beta \cos \theta \sin \varphi_k 
\\
-\beta \sin \theta
\end{pmatrix}
=
\\
= \frac{B_0}{2} e^{ik_z z + im \varphi} 
\begin{pmatrix}
-e^{i\varphi} J_{m+1}(\kappa \rho) o_- - e^{-i\varphi} J_{m-1}(\kappa \rho) o_+ \\
ie^{i\varphi}J_{m+1}(\kappa \rho) o_- - ie^{-i\varphi}J_{m-1}(\kappa \rho) o_+ \\
-2\beta \sin\theta J_m(\kappa \rho) 
\end{pmatrix},
\end{multline}
where $o_\pm = \alpha \pm i \beta \cos\theta$. It may be also useful to transform the above expression into the column vector written in cylindrical coordinate system ($\rho, \varphi, z$):
\begin{equation}
\label{HApp}
\bH^{(e)} =
\begin{pmatrix}
H_\rho^{(e)} \\
H_\varphi^{(e)} \\
H_z^{(e)}
\end{pmatrix}
=
\frac{B_0}{2} e^{ik_z z + im \varphi} 
\begin{pmatrix}
- J_{m+1}(\kappa \rho) o_- - J_{m-1}(\kappa \rho) o_+ \\
i J_{m+1}(\kappa \rho) o_- - i J_{m-1}(\kappa \rho) o_+ \\
-2\beta \sin\theta J_m(\kappa \rho) 
\end{pmatrix}.
\end{equation}
\end{widetext}

\end{document}